\documentstyle[aps,floats,amssymb,amsmath,bm,epsfig]{revtex}
\begin{document}
\bibliographystyle{prsty}

\title{Local Search in Unstructured Networks}
\author{Lada A. Adamic
\thanks{email: ladamic@hpl.hp.com}
\and
    Rajan M. Lukose
    \thanks{email: lukose@hpl.hp.com}
\and
    Bernardo A. Huberman
    \thanks{email: huberman@hpl.hp.com}
\\ HP Labs
\\ Palo Alto, CA 94304
}

\maketitle

\section{Introduction}

Recently, studies of networks in a wide variety of fields, from
biology to social science to computer science, have revealed some
commonalities \cite{albert02review}. It has become clear that the
simplest classical model of random networks, the Erdos-Renyi model
\cite{erdos60randgraph}, is inadequate for describing the topology
of many naturally occurring networks.  These diverse networks are
more accurately described by power-law or scale-free link
distributions. In these highly skewed distributions, the
probability that a node has $k$ links is approximately
proportional to $1/k^{\tau}$. The link graph of the World Wide Web
\cite{adamic99smallworld}, the Internet router backbone
\cite{faloutsos99topology}, certain representations of biological
pathways \cite{barabasi00metabolic}, and some social networks
\cite{aiello00massrandom,newman02social,liljeros01sex,barabasi99scaling},
each have approximately power-law distributions, in contrast to
the Poisson distribution consistent with the Erdos-Renyi random
graph model.

In addition to the characterization of the topological structure
of these networks, other important questions concerning the
growth, robustness, and dynamics on such networks have been
addressed.  For example, the question of what dynamical models of
graph growth tend to generate power-law networks has been
investigated
\cite{barabasi99scaling,adamic99nature,krapivsky01growing}, as
well as their robustness with respect to error and attack
\cite{barabasi00error}.

Another important dynamical question is the behavior of local
search strategies on networks.  Much of the recent work on
networks has been motivated by the "small world" phenomenon, in
which even very large networks (possibly possessing local
clustering or structure) have very short diameters.  Here the
diameter is defined as the average shortest path length between
the nodes in the network.  The existence of this phenomenon has
been demonstrated in different kinds of networks
\cite{newman01graphs}, and the property of short paths is
obviously important for dynamic models such as disease spreading
\cite{pastorsatorras01epidemic} and message passing between
arbitrary nodes in a network.

The classic social experiment of Milgram \cite{milgram67} found
that people could find a short chain of acquaintances in order to
pass a message to each other, a phenomenon often referred to as
"six degrees of separation". This result was surprising given that
most people's interactions tend to be tied to their local
communities, with relatively few longer range connections. Watts
and Strogatz \cite{watts98smallworld} revitalized interest in the
small-world problem by showing that even in highly structured and
clustered graphs, a few long range connections dramatically reduce
the average shortest path length between nodes.

It is however another question how exactly participants in a
Milgram-style experiment might find these short paths, since they
do not have global knowledge of the whole graph. That is, even if
short paths exist, how can one (approximately) find them using
local information? Kleinberg
\cite{kleinberg00nav,kleinberg01dynamics} considered this question
for a lattice topology with distance dependent shortcuts and found
an elegant characterization of the conditions under which it is
possible to pass messages efficiently.  Kleinberg assumed a very
structured topology and considered algorithms which use the
target's position on a regular 2-D lattice to direct the search.

While the question of local search in real social networks is an
intriguing one, it also relates in an interesting way to recent
developments in information technology.  The internet and the
World Wide Web have certainly had an impact on the way that
millions of people all over the world communicate, affecting the
structure and dynamics of what we think of as traditional social
networks \cite{wellman01networks}.  These ever more ubiquitous
technologies, wired and wireless, tend to make geography and
distance less relevant for communication between people.

But the relationship is also bi-directional.  A social network is
also a metaphor that is relevant for understanding popular
internet technologies such as peer-to-peer (p2p) file-sharing
networks. These networks share some of the topological features of
social networks. The Gnutella system connects users computers
directly with others to share files, without a central point of
coordination. In such networks, the name of the target file may be
known, but due to the network's ad hoc nature, until a real-time
search is performed the node holding the file is not known. In
order to find files on the system, peers pass messages along to
the other peers that they know of. In contrast to the scenario
considered by Kleinberg, there is no global information about the
position of the target, and hence it is not possible to determine
whether a step is a move towards or away from the target.

These networks, while not centrally planned in structure, grow
according to a simple self-organizing process.  Recent
measurements of Gnutella \cite{clip200bwbarrier} and simulated
Freenet networks \cite{hong01freenet} show that they have
power-law degree distributions. The resulting highly unstructured
networks need efficient search algorithms in order to function
well.  These algorithms should rely on local information in order
to avoid a dependence on a central point of failure, and to
accommodate their dynamic nature.

In this chapter, we will discuss a number of message-passing
algorithms that can be efficiently used to search through
power-law networks. We will discuss relevant work from both the
statistical physics community as well as the computer science
community.  Most of these algorithms are meant to be improvements
for peer-to-peer file sharing systems, and some may also shed some
light on how unstructured social networks with certain topologies
might function relatively efficiently with local information. Like
the networks that they are designed for, these algorithms are
completely decentralized, and they exploit the power-law link
distribution in the node degree. The algorithms use local
information such as the identities and connectedness of their
neighbors, and their neighbors' neighbors, but not the target's
global position. We demonstrate that some of these search
algorithms can work well on real Gnutella networks, scale
sub-linearly with the number of nodes, and may help reduce the
network search traffic that tends to cripple such networks.

The chapter is organized as follows.  Sections
\ref{plsearchanalytic}, \ref{plsearchsimulation},
\ref{poissoncomparison}, and \ref{gnutellasection} review results
from Adamic et al.\cite{adamic01plsearch} regarding localized
search. Sections \ref{plsearchanalytic} and
\ref{plsearchsimulation} present analytical and simulation
results, section \ref{poissoncomparison} compares search in
Poisson random graphs and section \ref{gnutellasection} describes
the application of the algorithms to Gnutella. Section
\ref{pathfindingsection} examines the length of the paths found in
search, section \ref{shortestpathsection} looks into shortest
paths in power-law graphs, while section
\ref{adaptivesearchsection} examines search strategies based on
information learned about the network, and section
\ref{conclusionsection} concludes.

\section{Search in power-law random graphs }
\label{plsearchanalytic}
\subsection{Intuition}

\begin{figure}[tbp]
\begin{center}
\includegraphics[scale=1]{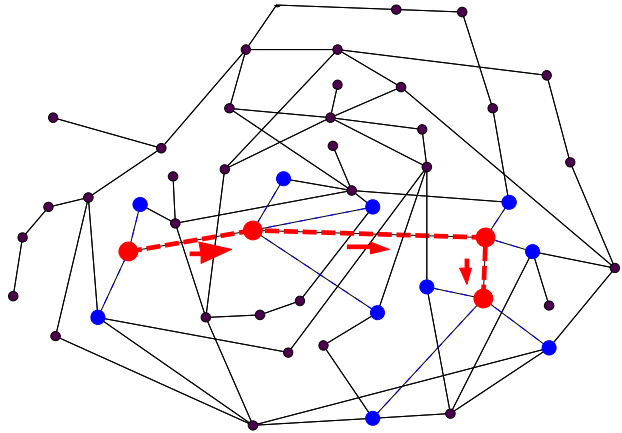}
\end{center}
\caption[An illustration of a search on a Poisson graph]{An
example of a search on a 50 nodes Poisson graph. Starting at a
node having 3 neighbors, the search finds 14 nodes in 4 steps.
\label{poissonwalk}}
\end{figure}

\begin{figure}[tbp]
\begin{center}
\includegraphics[scale=1]{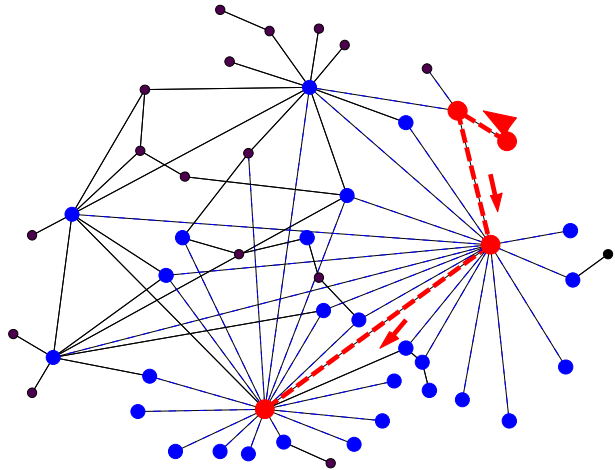}
\end{center}
\caption[An illustration of a search on a power-law graph]{An
illustration of a search on a power-law graph. Starting at a node
having a single neighbor, the search finds 31 nodes in 4 steps.
\label{plwalk}}
\end{figure}

The local search strategies we will be discussing use the
intuition that connections tend to be disproportionately
distributed among nodes and that the well-connected nodes should
provide access to a greater portion of the network. In figures
\ref{poissonwalk} and \ref{plwalk} we compare a sample walk on a
standard random graph with a Poisson degree distribution and a
power-law graph with the same number of nodes and edges. We plot
the number of nodes accessible as a message is passed through two
graphs, starting at a random node and proceeding toward the next
most highly connected neighbor. Since each node has knowledge of
its neighbors, we count reaching a node as finding all of its
previously undiscovered neighbors.

The search on the power-law graph finds 30 nodes in 4 steps, while
the same approach on the Poisson graph finds only 14 nodes in
spite of the initial node having higher degree. Even though the
two graphs have the same total number of edges, the distribution
of edges allows one to search the power-law graph more rapidly,
using only local information.

In following section we will follow up on this intuition and use
the generating function formalism introduced by Newman et al.
\cite{newman01graphs} for graphs with arbitrary degree
distributions to analytically characterize search-cost scaling in
graphs.

\subsection{Random walk search}
\label{randomwalksearch}

First we examine the number of nodes encountered in a random walk
on the graph. Let $G_0(x)$ be the generating function for the
distribution of the vertex degree $k$. Then
\begin{equation}
G_0(x) = \sum_{0}^{\infty}p_k x^k \label{G0defined}
\end{equation}
where $p_k$ is the probability that a randomly chosen vertex on
the graph has degree $k$.

For a graph with a power-law distribution with exponent $\tau$,
minimum degree $k = 1$ and an abrupt cutoff at $m = k_{max}$, the
generating function is given by

\begin{equation}
G_0(x) = c \sum_{1}^{m}k^{-\tau} x^k   \label{G0power}
\end{equation}
with $c$ a normalization constant which depends on $m$ and $\tau$
to satisfy the normalization requirement

\begin{equation}
G_0(1) = c \sum_{1}^{m} k^{-\tau} = 1
\end{equation}
The average degree of a randomly chosen vertex is given by

\begin{equation}
z_1 = <k> = \sum_{1}^{m} k p_{k} = G_0^{\prime}(1)
\label{firstneighbors}
\end{equation}
Note that the average degree of a vertex chosen at random and one
arrived at by following a random edge are different. A random edge
arrives at a vertex with probability proportional to the degree of
the vertex, i.e. $p^\prime(k) \sim k p_k$. The correctly
normalized distribution is given by

\begin{equation}
\frac{\sum_k k p_k x^k}{\sum_k k p_k} = x \frac{G^\prime_0
(x)}{G^\prime_0 (1) }
\end{equation}
If we want to count the number of outgoing edges from the vertex
we arrived at, but not include the edge we just came on, we need
to divide by one power of $x$. Hence the number of new neighbors
encountered on each step of a random walk is given by the
generating function

\begin{equation}
G_1(x) = \frac{G^\prime_0(x)}{G^\prime_0 (1)}
\end{equation}
where $G^\prime_0 (1)$ is the average degree of a randomly chosen
vertex as mentioned previously.

Since we are concerned with local search algorithms, we make the
reasonable assumption that nodes may have at least some knowledge
of their neighboring nodes' neighbors. Hence, we now compute the
distribution of second neighbors. The probability that any of the
2nd neighbors connects to any of the first neighbors or to one
another goes as $N^{-1}$ and can be ignored in the limit of large
$N$. Therefore, the distribution of the second neighbors of the
original randomly chosen vertex is determined by

\begin{equation}
\sum_k p_k[G_1(x)]^k = G_0(G_1(x))
\end{equation}
It follows that the average number of second neighbors is given by

\begin{equation}
z_{2A} = [\frac{\partial}{\partial x} G_0(G_1(x))]_{x=1} =
G^\prime_0 (1) G^\prime_1 (1) \label{2ndDegNeighRand}
\end{equation}
Similarly, if the original vertex was not chosen at random, but
arrived at by following a random edge, then the number of second
neighbors would be given by

\begin{equation}
z_{2B} = [\frac{\partial}{\partial x} G_1(G_1(x))]_{x=1} =
[G^\prime_1 (1)]^2 \label{2ndDegNeighFollow}
\end{equation}
In both Equation \ref{2ndDegNeighRand} and Equation
\ref{2ndDegNeighFollow}, the fact that $G_1 (1) = 1$ was used.
Both these expressions depend on the values $G^\prime_0 (1)$ and
$G^\prime_1 (1)$ so we calculate those for given $\tau$ and $m$.
For simplicity and relevance to most real-world networks of
interest we assume $2 < \tau < 3$.

\begin{equation}
G^\prime_0 (1) = \sum_1^m c k^{1-\tau} \sim \int_1^m x^{\tau -1}
dx = \frac{1}{\tau-2}(1 - m^{2-\tau})
\end{equation}

\begin{eqnarray}
G^\prime_1 (1) & = &\frac{1}{G^\prime_0 (1)}
\frac{\partial}{\partial x} \sum_1^m c k^{1-\tau} x^{k-1}
\\ & = & \frac{1}{G^\prime_0 (1)}\sum_2^m c k^{1-\tau} (k-1) x^{k-2}
\\ & \sim & \frac{1}{G^\prime_0 (1)} \frac{m^{3-\tau} (\tau - 2) -
2^{2-\tau} (\tau - 1) + m^{2 - \tau} (3 - \tau)}{(\tau-2)(3-\tau)}
\end{eqnarray}
for large cutoff values $m$. Now we impose the cutoff of Aiello et
al. \cite{aiello00massrandom} at $m \sim N^{1/\tau}$. The cutoff
is chosen so that in an non-truncated distribution the expected
number of nodes among $N$ having exactly the cutoff degree is 1.
No nodes of degree higher than the cutoff are present in the
graph. In real world graphs one does frequently observe nodes of
degree higher than this imposed cutoff, so that our calculations
describe a worse case scenario. Since $m$ scales with the size of
the graph $N$ and for $2 < \tau < 3$ the exponent $2-\tau$ is
negative, we can neglect terms constant in $m$. This leaves

\begin{equation}
G^\prime_1 (1)  = \frac{1}{G^\prime_0
(1)}\frac{m^{3-\tau}}{(3-\tau)}
\end{equation}
Substituting into Equation \ref{2ndDegNeighRand} (the starting
node is chosen at random) we obtain
\begin{equation}
z_{2A} = G^\prime_0 (1) G^\prime_1 (1) \sim m^{3-\tau}
\end{equation}

We can also derive $z_{2B}$, the number of 2nd neighbors
encountered as one is doing a random walk on the graph.

\begin{equation}
z_{2B} = [G^\prime_1 (1)]^2 =
[\frac{\tau-2}{1-m^{2-\tau}}\frac{m^{3-\tau}}{3-\tau}]^2
\label{G12eq}
\end{equation}
Letting $m \sim N^{1/\tau}$ as above, we obtain

\begin{equation}
z_{2B} \sim N^{2 (\frac{3}{\tau} - 1)}
\end{equation}
Thus, as the random walk along edges proceeds node to node, each
node reveals more of the graph since it has information not only
about itself, but also of its neighborhood.   The search cost $s$
is defined as the number of steps until approximately the whole
graph is revealed so that $s \sim N/z_{2B}$, or

\begin{equation}
s \sim N^{3(1-2/\tau)} \label{RandStratScale}
\end{equation}
In the limit $\tau \rightarrow 2$, equation \ref{G12eq} becomes

\begin{equation}
z_{2B} \sim \frac{N}{ln^2(N)}
\end{equation}
and the scaling of the number of steps required is

\begin{equation}
s \sim ln^2(N)
\end{equation}

\subsection{Search utilizing high degree
nodes}
\label{highdegreesearch}

Random walks in power-law networks naturally gravitate towards the
high degree nodes, but an even better scaling is achieved by
intentionally selecting high degree nodes. For $\tau$ sufficiently
close to $2$ one can approximately walk down the degree sequence,
visiting the node with the highest degree, followed by a node of
the next highest degree, etc. Let $m - a$ be the degree of the
last node we need to visit in order to scan a certain fraction of
the graph. We make the self-consistent assumption that $a << m$,
i.e. the degree of the node has not dropped too much by the time
we have scanned a fraction of the graph. Then the number of first
neighbors scanned is given by
\begin{figure}[tbp]
\begin{center}
\includegraphics[scale=0.5]{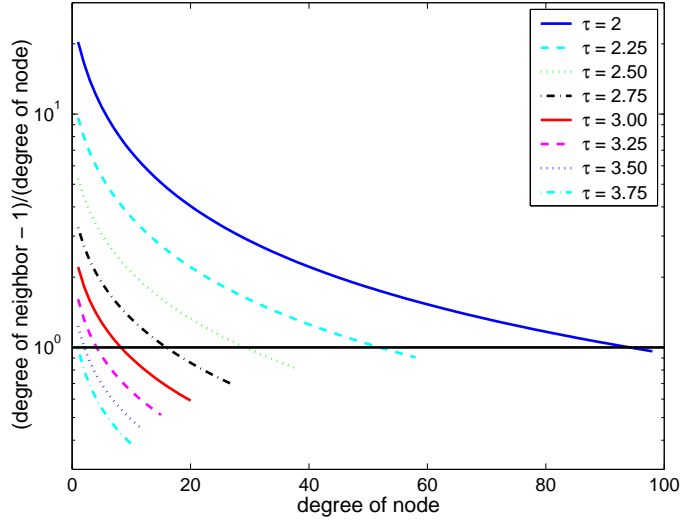}
\end{center}
\caption[Ratio of richest neighbor and the node itself for various
power-law exponents]{Ratio $r$ (the expected degree of the richest
neighbor of a node whose degree is $n$ divided by $n$) vs. $n$ for
$\tau$ (top to bottom) $=$ 2.0, 2.25, 2.5, 2.75, 3.00, 3.25, 3.50,
and 3.75. Each curve extends to the cutoff imposed for a 10,000
node graph with the particular exponent.\label{degreeratios}}
\end{figure}

\begin{equation}
z_{1D} = \int^{m}_{m-a} N k^{1-\tau} dk \sim N a m^{1-\tau}
\end{equation}
The number of nodes having degree between $m-a$ and $m$, or
equivalently, the number of steps taken is given by
$\int^{m}_{m-a} k^{-\tau} \sim a$. The number of second neighbors
when one follows the degree sequence is given by:

\begin{equation}
z_{1D} \ast G^{\prime}_{1}(1) \sim N a m^{2(2-\tau)}
\end{equation}
which gives the number of steps required as

\begin{equation}
s \sim m^{2(\tau-2)} \sim N^{2-\frac{4}{\tau}}
\label{DegStratScale}
\end{equation}

We now consider when and why it is possible to go down the degree
sequence. We start with the fact that the original degree
distribution is a power-law:
\begin{equation}
p(x) = (\sum_{1}^{m} x^{-\tau})^{-1} x^{-\tau}
\end{equation}
where $m=N^{1/\tau}$ is the maximum degree. A node chosen by
following a random link in the graph will have its remaining
outgoing edges distributed according to

\begin{equation}
p^\prime(x) = \left[\sum_{0}^{m-1}
(x+1)^{(1-\tau)}\right]^{-1}(x+1)^{(1-\tau)}
\end{equation}
At each step one can choose the highest degree node among the $n$
neighbors. The expected number of the outgoing edges of that node
can be computed as follows. In general, the cumulative
distribution (CDF) \(P_{max}(x,n) \) of the maximum of $n$ random
variables can be expressed in terms of the CDF \( P(x) = \int_0^x
p(x\prime)\, \mathrm{d}x\prime \) of those random variables:
\(P_{max}(x,n)=P(x)^n \). This yields
\begin{equation}
p^\prime_{max}(x,n) = n (1+x)^{1-\tau} (\tau - 2) \left[1 -
(x+1)^{2 - \tau}\right]^{n-1} (1 - N^{2/\tau - 1})^{-n}
\label{PMaxDist}
\end{equation}
for the distribution of the number of links the richest neighbor
among $n$ neighbors has.

Finally, the expected degree of the richest node among n is given
by
\begin{equation}
E[x_{max}(n)]=\sum^{m-1}_0 x p^\prime_{max}(x,n)
\end{equation}
Numerically integrating the above equation yields the ratio
between the degree of a node and the expected degree of its
richest neighbor, plotted in Figure \ref{degreeratios}. For a
range of exponents and node degrees, the expected degree of the
richest neighbor is higher than the degree of the node itself.
However, as one moves to nodes of higher and higher degree, the
probability of finding a neighbor with an even higher degree
starts falling (the precise point depends strongly on the
power-law exponent).

What this means is that one can approximately follow the degree
sequence across the entire graph for a sufficiently small graph or
one with a power-law exponent close to 2 ($ 2.0 < \tau < 2.3$). At
each step one chooses a node with degree higher than the current
node, quickly finding the one with the highest degree. Once the
highest degree node has been visited, it will be avoided, and a
node of approximately second highest degree will be chosen.
Effectively, after a short initial climb, one goes down the degree
sequence. This is the most efficient way to do this kind of
sequential search.

\section{Simulation}
\label{plsearchsimulation}
We used simulations on a random network
with a $\tau = 2.1$ power-law link distribution and a simple
cutoff at $m \sim N^{1/\tau}$ to validate our analytical results.
The graph is generated by assigning links at random between nodes
of pre-assigned degree drawn from the power-law distribution. For
$2 < \tau < 3.48$, a graph contains a giant connected component
(GCC), the largest group of nodes such that any node can be
reached from any other node following links
\cite{aiello00massrandom}. All our measurements were performed on
the GCC which contained the majority of the nodes of the original
graph and most of the links as well. The link distribution of the
GCC is nearly identical to that of the original graph with a
slightly smaller number of nodes of degree 1 and 2.

Next we apply our message passing algorithm to the network. Two
nodes, the source and the target, are selected at random. At each
time step the node which has the message passes it on to one of
its neighbors. The process ends when the message is passed on to a
neighbor of the target, that, knowing the identity of its
neighbors, passes the message to the target directly. The process
is analogous to performing a random walk on a graph, where each
node is 'visited' as it receives the message.

There are several variants of the algorithm, depending on the
strategy and the amount of local information available.
\begin{enumerate}
\item The node passes the message on to one of its neighbors at
random, optionally avoiding a node which has already seen the
message.

\item The node knows the degrees of its neighboring nodes and chooses to
pass the message onto the neighbor with the most neighbors.

\item The node knows who its neighbors' neighbors are and passes the message onto a
neighbor of the target if possible.
\end{enumerate}

In order to avoid passing the message to a node that has already
seen the message, the message itself must be signed by the nodes
as they receive the message. Further, if a node has passed the
message, and finds that all of its neighbors are already on the
list, it puts a special mark next to its name, which means that it
is unable to pass the message onto any new node. This is
equivalent to marking nodes as follows:

\begin{description}
\item[white] Node has not been visited.
\item[gray] Node has been visited, but all its neighbors have not.
\item[black] Node and all its neighbors have been visited already.
\end{description}

Here we compare two strategies. The first performs a random walk,
where only retracing the last step is disallowed. In the message
passing scenario, this means that if Bob just received a message
from Jane, he wouldn't return the message to Jane if he could pass
it to someone else. The second strategy is a self avoiding walk
which avoids passing the message to previously visited nodes and
prefers high degree nodes to low degree ones. In both strategies
the first and second neighbors are scanned at each step.

\begin{figure}[tbp]
\begin{center}
\includegraphics[scale=0.7]{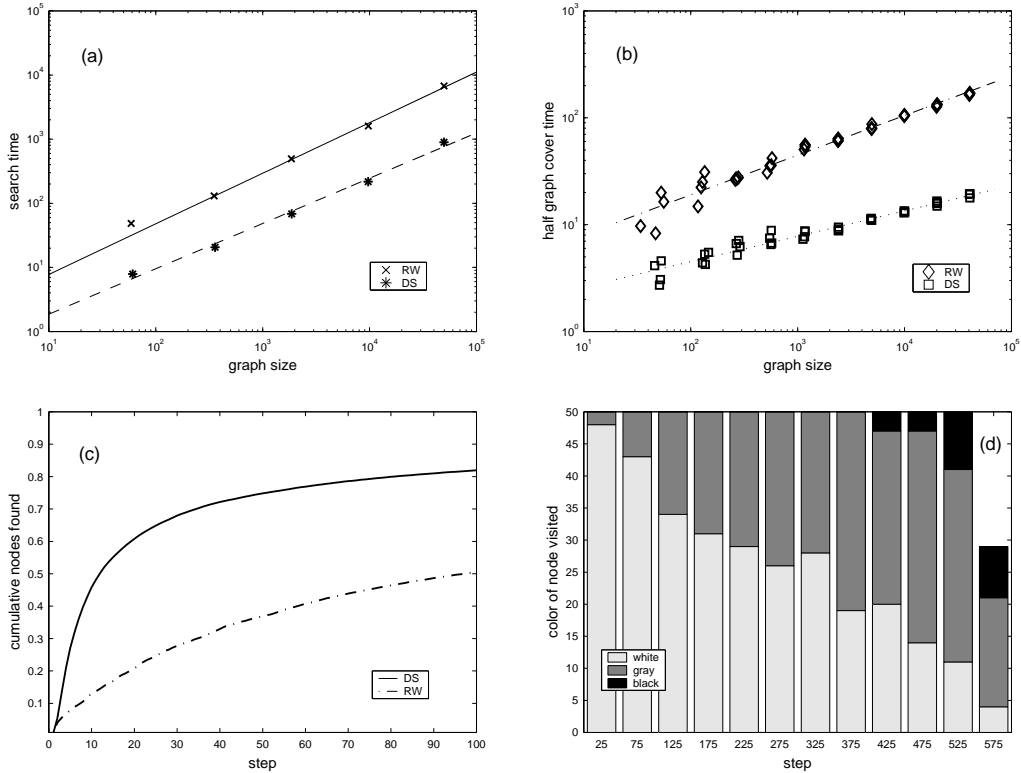}
\end{center}
\caption[Simulated search in power-law graphs]{(a) Scaling of the
average node-to-node search cost in a random power-law graph with
exponent 2.1, for random walk (RW) and high-degree seeking (DS)
strategies. The solid line is a fitted scaling exponent of 0.79
for the RW strategy and the dashed is an exponent of 0.70 for the
DS strategy. (b) The observed and fitted scaling for half graph
cover times for the RW and DS strategies. The fits are to scaling
exponents of 0.37 and 0.24, respectively. (c) Cumulative
distribution of nodes seen vs the number of steps taken for the RW
and DS strategies on a 10 000 node graph. (d) Bar graph of the
color of nodes visited in DS search of a random 1000 node
power-law graph with exponent 2.1. White represents a fresh node,
gray represents a previously visited node that has some unvisited
neighbors, and black represents nodes for which all neighbors have
been previously visited.} \label{fourinone}
\end{figure}

Figure \ref{fourinone}(a) shows the scaling of the average search
time with the size of the graph for the two strategies. The
scaling (exponent 0.79 for the random walk and 0.70 for the high
degree strategy) is not as favorable as in the analytic results
derived above ($0.14$ for the random walk and $0.1$ for the high
degree strategy when $\tau = 2.1$) .

Consider, on the other hand, the number of steps it takes to cover
half the graph. That is, instead of asking how long it would take
on average to find any node in the graph, we ask how long it would
take to find the first $50\%$ of the nodes. Such a measure is
reasonable in a network where more than one node is likely to be
able to satisfy a request. In a social context, there might be
more than one person who has a particular item or can share
expertise on a subject. In the context of a file sharing network,
there might be more than one node having the requested file.

For this measure we observe a scaling which is much closer to the
ideal. As shown in Figure \ref{fourinone}(b), the cover time
scales as $N^{0.37}$ for the random walk strategy vs. $N^{0.15}$
from Equation \ref{RandStratScale}. Similarly, the high degree
strategy cover time scales as $N^{0.24}$ vs. $N^{0.1}$ in Equation
\ref{DegStratScale}.

The difference in the value of the scaling exponents of the cover
time and average search time implies that a majority of nodes can
be found fairly efficiently, but others demand high search costs.
As Figure \ref{fourinone}(c) shows, a large portion of the 10,000
node graph is covered within the first few steps, but some nodes
take as many steps or more to find as there are nodes in total.
For example, the high degree seeking strategy finds about 50\% of
the nodes within the first 10 steps (meaning that it would take
about  $10 + 2 = 12$ hops to reach 50\% of the graph). However,
the skewness of the search time distribution bring the average
number of steps needed to 217.

Some nodes take a long time to find because the random walk, after
a brief initial period of exploring fresh nodes, tends to revisit
nodes. It is a well-known result that the stationary distribution
of a random walk on an undirected graph is simply proportional to
the distribution of links emanating from a node. Thus, nodes with
high-degree are often revisited in a walk.

A high-degree seeking self-avoiding walk is an improvement over
the random walk taking 13 times fewer steps, but still cannot
avoid retracing its steps. Figure \ref{fourinone}(d) shows the
color of nodes visited on such a walk for a $N = 1000$ node
power-law graph with exponent 2.1 and an abrupt cutoff at
$N^{1/2.1}$. The number of nodes of each color encountered in 50
step segments is recorded in the bar for that time period, showing
that some grey and black nodes were encountered before the all of
the nodes were found.

Although revisiting nodes slows down search, it is the form of the
link distribution that is responsible for changes in the search
cost scaling. In a graph with a uniform link distribution the
number of new nodes discovered at every step would be proportional
to the number of unexplored nodes in the graph. The factor by
which the search is slowed down through revisits would be
independent of the size of the graph.

In contrast, in a power-law graph, a large number of links point
to only a small subset of high degree nodes. When a new node is
visited, its links do not let us uniformly sample the graph, they
preferentially lead to high degree nodes, which have likely been
seen or visited in a previous step. Ironically, the presence of
high degree nodes, so useful to our search strategies, also
worsens the search cost scaling from the ideal scaling found in
sections \ref{randomwalksearch} and \ref{highdegreesearch}. This
would not be true of a Poisson graph, where all the links are
randomly distributed and hence all nodes have approximately the
same degree. We will explore and contrast the search algorithm on
a Poisson graph in the following section.

\section{Comparison with Poisson distributed graphs}
\label{poissoncomparison}
\begin{figure}[tbp]
\begin{center}
\includegraphics[scale=0.5]{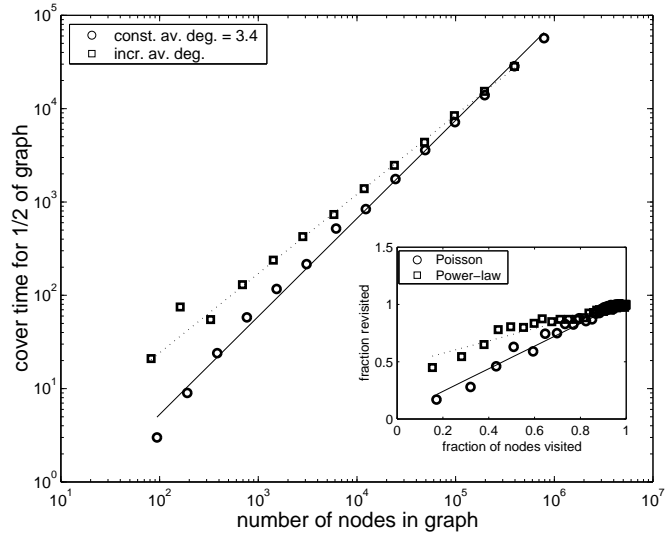}
\end{center}
\caption[Search time scaling in a Poisson graph]{Squares are
scaling of cover time for 1/2 of the graph for a Poisson graph
with a constant average degree/node (with fit to a scaling
exponent of $1.0$). Circles are the scaling for Poisson graphs
with the same average degree/node as a power-law graph with
exponent 2.1 (with fit to a scaling exponent of $0.85$).  The
inset compares revisitation between search on Poisson versus
power-law graphs, as discussed in the text. }
\label{poissonscaling}
\end{figure}

In a Poisson random graph with $N$ nodes
and $z$ edges, the probability $p = z/N$ of an edge between any
two nodes is the same for all nodes. The generating function
$G_0(x)$ is given by \cite{newman01graphs}:
\begin{equation}
G_0(x) = e^{z(x-1)}
\end{equation}
In this special case $G_0(x) = G_1(x)$, so that the distribution
of outgoing edges of a node is the same whether one arrives at the
vertex by following a link or picks the node at random. This makes
the analysis of search in a Poisson random graph particularly
simple. The expected number of new links encountered at each step
is a constant $p$, so that the number of steps needed to cover a
fraction $c$ of the graph is $s = c N/p$. If $p$ remains constant
as the size of the graph increases, the cover time scales linearly
with the size of the graph. This has been verified via simulation
of the random walk search as shown in Figure \ref{poissonscaling}.

In our simulations, in order to keep the total number of edges
equal between power-law and Poisson graphs of the same size, the
probability $p$ increases with the size. It grows slowly towards
its asymptotic value because of the particular choice of cutoff at
$m \sim N^{(1/\tau)}$ for the power-law link distribution. We
generated Poisson graphs with the same number of nodes and links
for comparison. Within this range of graph sizes, growth in the
average number of links per node appears as $N^{0.6}$, making the
average number of 2nd neighbors scale as $N^{0.15}$. This means
that the scaling of the cover time scales as $N^{0.85}$, as shown
in Figure \ref{poissonscaling}.

Note how well the simulation results match the analytical
expression. This is because nodes can be sampled in an
approximately even fashion by following links as is illustrated in
Figure \ref{poissonscaling}(inset). If links are evenly
distributed among the nodes, then when the search has covered 50\%
of the graph, one would expect to revisit previously seen nodes
about 50\% of the time. This is indeed the case for the Poisson
graph.

However, for the power-law graph, when 50\% of the graph has been
visited, nodes are revisited about 80\% of the time, which implies
that the same high degree nodes are being revisited before new
low-degree ones. This bias introduces a discrepancy between the
analytic scaling and the simulated results in the power-law case.
\begin{figure}[tbp]
\begin{center}
\includegraphics[scale=0.5]{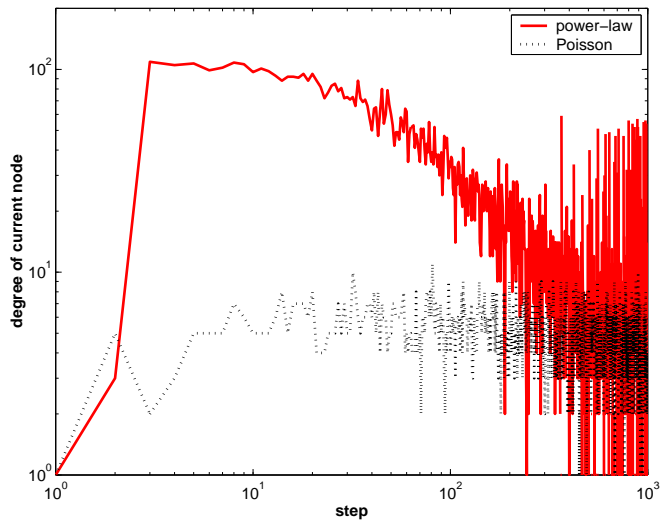}
\end{center}
\caption[Degrees of nodes visited in search]{Degrees of nodes
visited in a single search for power-law and Poisson graphs of
10,000 nodes.} \label{DegSequence}
\end{figure}
However, even the simulated $N^{0.35}$ scaling for a random,
minimally self-avoiding strategy on the power-law graph
out-performs the ideal $N^{0.85}$ scaling for the Poisson graph.
It's also important to note that the the high degree node seeking
strategy has a much greater success in the power-law graph because
it relies heavily on the fact that the number of links per node
varies considerably from node to node. In the Poisson graph, the
variance in the number of links is much smaller, making the high
degree node seeking strategy comparatively ineffective as shown
in.

Figure \ref{DegSequence} shows an illustration of this point. We
repeat the experiment in Figures \ref{poissonwalk} and
\ref{plwalk} on larger power-law and Poisson graphs with $N =
10,000$. In the power-law graph we start from a randomly chosen
node. In this case the starting node has only one link, but two
steps later we find ourselves at a node with the highest degree.
From there, one approximately follows the degree sequence, that
is, the node richest in links, followed by the second richest
node, etc. The strategy has allowed us to scan the maximum number
of nodes in the minimum number of steps. In comparison, the
maximum degree node of the exponential graph is 11, and it is
reached only on the 81st step. Even though the two graphs have a
comparable number of nodes and edges, the exponential graph does
not lend itself to quick search.

\section{Gnutella}
\label{gnutellasection}
\begin{figure}[tbp]
\begin{center}
\includegraphics[scale=0.70]{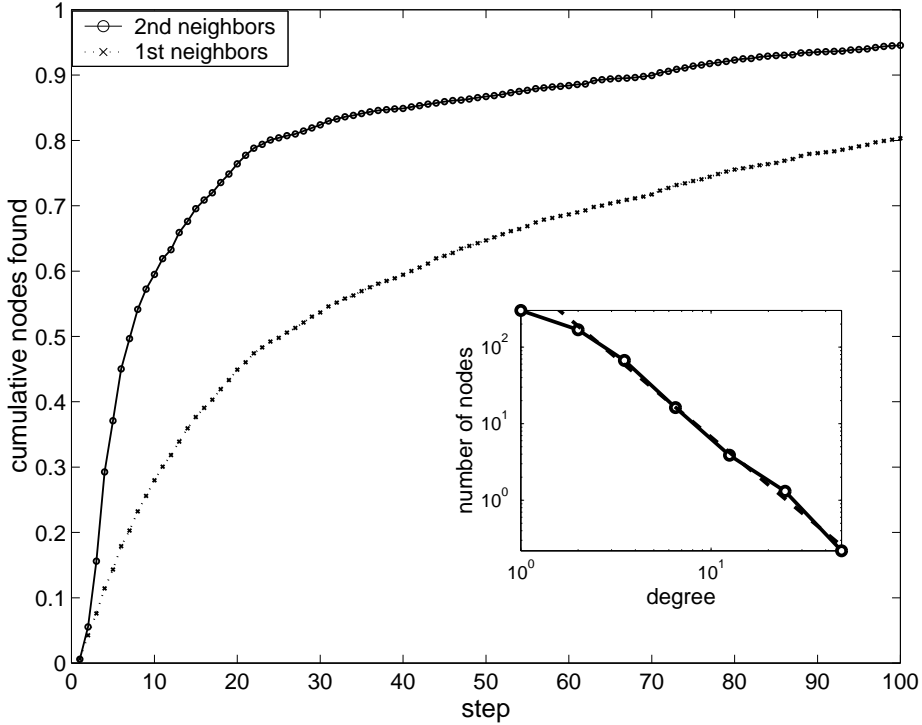}
\end{center}
\caption[Cumulative number of nodes found at each step in the
Gnutella network]{Cumulative number of nodes found at each step in
the Gnutella network.} \label{gnutellapass}
\end{figure}

Gnutella is a peer-to-peer filesharing system which treats all
client nodes as functionally equivalent and lacks a central server
which can store file location information. This is advantageous
because it presents no central point of failure. The obvious
disadvantage is that the location of files is unknown. When a user
wants to download a file, she sends a query to all the nodes
within a neighborhood of size $ttl$, the time to live assigned to
the query. Every node passes on the query to all of its neighbors
and decrements the $ttl$ by one. In this way, all nodes within a
given radius of the requesting node will be queried for the file,
and those who have matching files will send back positive answers.

This broadcast method will find the target file quickly, given
that it is located within a radius of $ttl$.  However,
broadcasting is extremely costly in terms of bandwidth. Every node
must process queries of all the nodes within a given $ttl$ radius.
In essence, if one wants to query a constant fraction of the
network, say 50\%, as the network grows, each node and network
edge will be handling query traffic which is proportional to the
total number of nodes in the network.

Such a search strategy does not scale well. As query traffic
increases linearly with the size of Gnutella graph, nodes become
overloaded as was shown in a study by Clip2
\cite{clip200bwbarrier}. 56k modems are unable to handle more than
20 queries a second, a threshold easily exceeded by a network of
about 1,000 nodes. With the 56k nodes failing, the network becomes
fragmented, allowing users to query only small section of the
network.

The search algorithms described in the previous sections may help
ameliorate this problem. Instead of broadcasting a query to a
large fraction of the network, a query is only passed onto one
node at each step.  The search algorithms are likely to be
effective because the Gnutella network has a power-law
connectivity distribution as shown in the inset of Figure
\ref{gnutellapass}.

Typically, a Gnutella client wishing to join the network must find
the IP address of an initial node to connect to. Currently, ad hoc
lists of "good" Gnutella clients exist \cite{clip200bwbarrier}. It
is reasonable to suppose that this ad hoc method of growth would
bias new nodes to connect preferentially to nodes which are
already fairly well-connected, since these nodes are more likely
to be "well-known".  Based on models of graph growth
\cite{barabasi99scaling,adamic99nature} where the "rich get
richer", the power-law connectivity of ad hoc peer-to-peer
networks may be a fairly general topological feature.

By passing the query to every single node in the network, the
Gnutella algorithm fails to take advantage of its connectivity
distribution. To implement our algorithm the Gnutella clients must
be modified to keep lists of the files stored by their first and
second degree neighbors have. This information must be passed at
least once when a new node joins the network, and it may be
necessary to periodically update the information depending on the
typical lifetime of nodes in the network. The importance of
localized indexing to scalability has been illustrated by the
growth of the FastTrack \cite{truelove01morpheus} network whose
size has reached hundreds of thousands of nodes. FastTrack is a
network similar to Gnutella, with no central server, but using
local indexing. A fraction of FastTrack clients with high
bandwidth and reliability are selected to be supernodes.
Supernodes index the files of other nodes and route queries on
their behalf. We note that unlike FastTrack, our algorithm
requires each node to store a local index.

Keeping track of the filenames of its neighbors' files places an
additional cost on every node. Since network connections saturated
by query traffic are a major weakness in Gnutella, and since
computational and storage resources are likely to remain much less
expensive than bandwidth, such a tradeoff is readily made.
However, now instead of every node having to handle every query,
queries are routed only through high connectivity nodes, a
situation similar to that of supernodes in the FastTrack network.
Since nodes can select the number of connections that they allow,
high degree nodes are presumably high bandwidth nodes that can
handle the query traffic. The network has in effect created local
directories valid within a two link radius. It is resilient to
attack because of the lack of a central server. As for power-law
networks in general \cite{barabasi00error}, the network is more
resilient than random graphs to random node failure, but less
resilient to attacks on the high degree nodes.

Further adjustments to the present Gnutella clients to implement
our algorithm involve switching from broadcasting queries to
passing them only to the highest degree nodes. To execute a
self-avoiding search, nodes need to append their IDs to the query
as they process it.

Figure \ref{gnutellapass} shows the success of the high degree
seeking algorithm on the Gnutella network. We simulated the search
algorithm on a crawl by Clip2 of the actual Gnutella network of
approximately 700 nodes. Assuming that every file is stored on
only one node,  50\% of the files can be found in 8 steps or less.
Furthermore, if the file one is seeking is present on multiple
nodes, the search will be even faster.

To summarize, we have argued that truly peer-to-peer networks like
Gnutella are likely to have a power-law structure, and that the
local search algorithms we have described can be effective.  As
the number of nodes increases, the (already small) number of nodes
that will need to be queried increases sub-linearly. As long as
the high degree nodes are able to carry the traffic, the Gnutella
network's performance and scalability may improve by using these
search strategies.

\section{Path finding}
\label{pathfindingsection}
So far we have only discussed the
amount of time it takes to locate a node a single time. But in the
process of searching for a node, one is also mapping out a path
which could be used to contact that node in the future. Removing
loops and backtracking steps from the search path leaves a route
to the desired node. This route could be reused should one desire
to communicate with the node again.

Kim et al. \cite{kim02pathfinding} have shown that following a
high-degree seeking strategy on power-law graphs produces paths
which scale on average as the logarithm of the size of the
network. While the paths found are not always the shortest paths
themselves, they share in the logarithmic scaling of the average
shortest path. In contrast, random walker strategies, or
strategies on non-power-law graphs such as Poisson random graphs
of small world graphs defined by Watts and Strogatz
\cite{watts98smallworld}, produce paths which whose scaling is
power-law.

\begin{figure}[tbp]
\begin{center}
\includegraphics[scale=0.65]{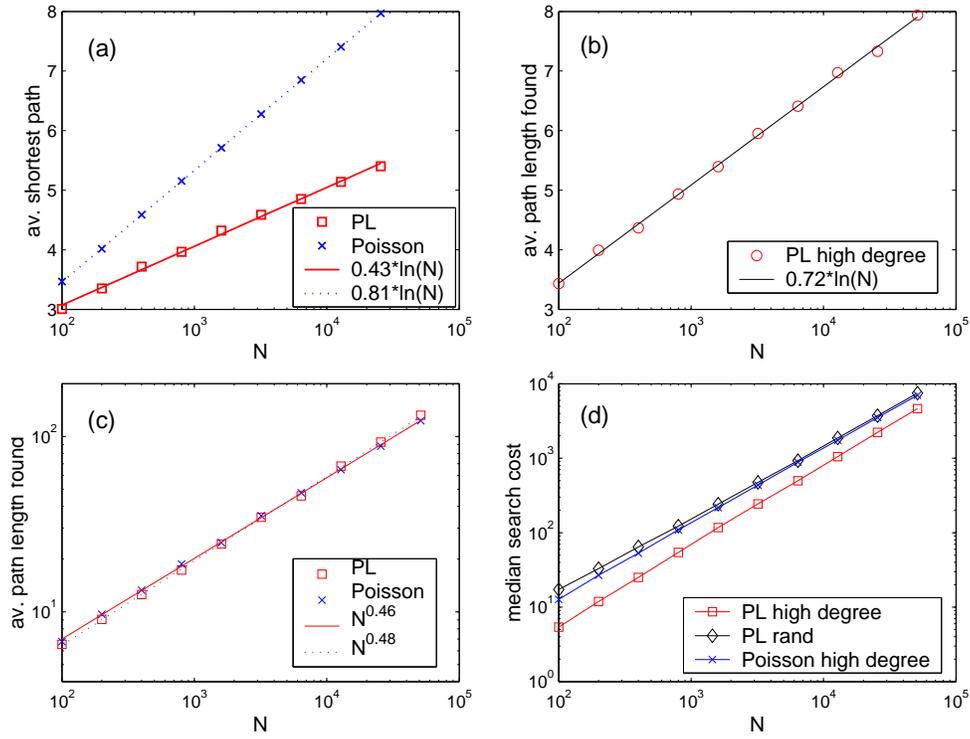}
\end{center}
\caption[Scaling of path finding strategies]{Scaling of path
finding strategies: (a) average shortest path found using breadth
first search for Poisson and power-law graphs, (b) average path
length found using a high degree strategy on a power-law graph,
(c) average path length found using a random strategy on a
power-law graph and a high degree strategy on a Poisson graph, (d)
median number of steps required to find a path between two nodes}
\label{pathfinding}
\end{figure}

Following the methods of Kim et al, we constructed a scale-free
network of  Barabasi and Albert (BA) \cite{barabasi99scaling}
type. Starting with a small number ($m_{0} = 2$) of vertices, a
new vertex with $m = 2$ edges is added at each time step such that
the probability of an edge connecting to a vertex is proportional
to the degree of the vertex. This method yields a power-law
network which has an exponent $\tau = 3$ but is not truly random
because correlations between node degrees do exist. Although
$\tau=3$ lies outside the regime favorable to the previously
discussed search strategies, requiring many steps to locate a
node, the paths obtained with the loops removed scale
logarithmically with the size of the network.

Figure \ref{pathfinding} shows a comparison between the actual
shortest paths and the shortest paths found using various search
strategies on BA power-law graphs and Poisson graphs with the same
total number of vertices and edges. Figure \ref{pathfinding}a
shows the that shortest paths scale logarithmically in the size of
the graph in both power-law and Poisson graphs, but the average
shortest path in a power-law graph grows more slowly as the number
of nodes increases. In effect, high degree nodes are drawing the
graph closer together. This will be discussed further in section
\ref{shortestpathsection}.

In order to find the exact shortest path, a broadcast method
equivalent to a breadth first search (BFS) must be used. As
mentioned in our discussion of Gnutella, broadcasting can
overwhelm the bandwidth resources of the network. Kim et al.
propose instead a search similar to the strategies discussed in
sections \ref{randomwalksearch} and \ref{highdegreesearch}. The
message is passed from only one node at each step, either
randomly, or to the highest degree neighbor, using knowledge only
of the first degree neighbors and their degree. When following the
high-degree strategy, a node passes the message to the highest
degree node it personally has not passed the message to
previously.

This strategy is not truly self-avoiding in the sense that a node
does not try to avoid passing the message onto a node that others
have already contacted. Curiously, we find that a truly
self-avoiding strategy, while locating nodes more quickly, does
not yield short paths in the end. Kim et al. also note that if the
strategy chooses nodes probabilistically, with the probability of
a node being chosen proportional to its degree, the logarithmic
scaling is lost. It is possible that both the self-avoiding and
probabilistic methods fare worse because they return to the higher
degree nodes less frequently. Because the majority of paths pass
through high-degree nodes, the deterministic strategy which
routinely revisits high degree nodes before moving forward is more
likely to find a shorter path.

For comparison, we also plot in Figure \ref{pathfinding}c the
length of the shortest paths found in the BA graph by choosing
nodes at random rather than based on their connectivity. The paths
found have a much less favorable power-law scaling of
approximately $N^{0.5}$, compared to the logarithmic scaling of
the shortest path. A similar result is obtained when using a high
degree strategy on an equivalent Poisson graph, where extremely
high degree nodes are absent.

Even in the case where short paths can be found using a high
degree strategy on a power-law graph, the approach may be too
costly. While the length of the average path found grows slowly as
the size of the network increases, the average cost in the amount
of time necessary to find the path, shown in Figure
\ref{pathfinding}d, scales nearly linearly. The median number of
steps required to find a node grows into the thousands while
targets remain less than 10 steps away.

Although the above discussion of path finding strategies
demonstrates how nodes could in principle find shortest paths
between each other, the extremely high cost of this procedure
suggests that additional clues as to the location of the target or
knowledge of second degree neighbors would be necessary to make
such an approach worthwhile.

\section{Shortening the shortest path}
\label{shortestpathsection}
The previous sections described the
role high degree nodes play in locating nodes and constructing a
short path to a target. A further twist however, is the fact that
the presence of high degree nodes shortens the shortest paths
themselves. The average shortest path grows more slowly as the
size of the network increases in a power-law graph than in an
equivalent Poisson random graph.

\begin{figure}[tbp]
\begin{center}
\includegraphics[scale=0.5]{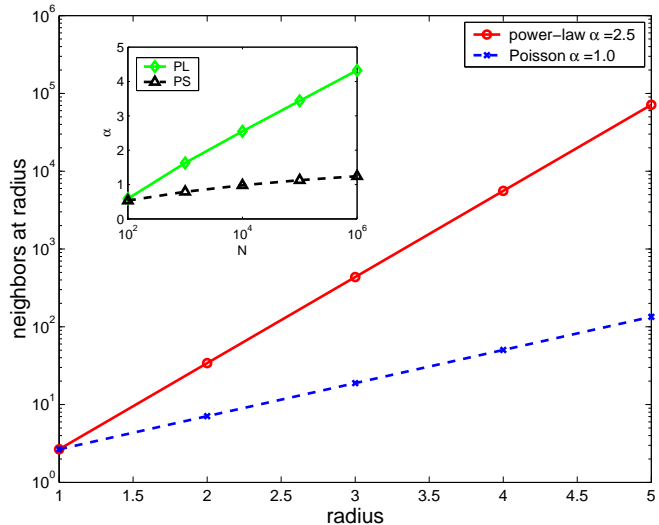}
\end{center}
\caption[Number of neighbors vs. radius]{The expected number of
nodes at a given distance from a node is plotted for $\tau = 2.1$
power-law and Poisson graphs with 10,000 nodes and the same number
of edges. The number of neighbors as a function of the radius $r$
is $\exp(\alpha*r)$. $\alpha=2.5$ for the power-law graph, while
$\alpha=1.0$ for the Poisson one. The actual number of neighbors
at higher radii is limited by the finite size of the graphs. The
inset shows the variation of $\alpha$ with the size of the
network. } \label{neighborsandradius}
\end{figure}
Figure \ref{neighborsandradius} shows the number of neighbors who
are $r$ steps away from a randomly chosen vertex given by the
formula of Newman et al. \cite{newman01graphs}:

\begin{equation}
z_r = [\frac{z_{2A}}{z_{1}}]^{r-1} z_{1}  \label{rthneighbors}
\end{equation}
with $z_{1}$ and $z_{2A}$ given by Equations \ref{firstneighbors}
and \ref{2ndDegNeighRand}. By choice, the Poisson graph has the
same average number $z_{1}$ of first degree neighbors. Using the
above result that the expected number of outgoing edges following
a link is equal to the average vertex degree, the number of second
degree neighbors is simply $z_{1}^2$.

The number of nodes at distance $r$ scales as $\exp(\alpha r)$,
$\alpha = z_{2A}/z_{1}$. The actual number of neighbors of course
cannot continue to increase exponentially with $r$ due to the
finite size of the graph. For a Poisson graph $\alpha = z_{1}$,
where $z_{1}$ is determined by the degree distribution of the
power-law graph and asymptotes as the size of the graph increases.
The ratio of the number of second degree to first degree neighbors
grows more rapidly for a power-law graph as shown in the inset of
Figure \ref{neighborsandradius}.

\subsection{Iterative deepening}
\label{iterativedeepening}

The fact that the number of hops between nodes is shorter in a
power-law graph implies that the broadcasting method of locating
nodes and resources will return results more quickly. This is
because, as shown in Figure \ref{neighborsandradius}, compared to
other graph topologies, many more nodes are available at the same
radius.   Yang and Garcia-Molina \cite{yang02improvingsearch} and
Lv et al. \cite{lv02ptopsearch} have experimented with a local
search method that benefits from this fact in order to improve
upon the standard fixed-radius broadcast that the default Gnutella
protocol uses.

Yang and Garcia-Molina's method, which they call iterative
deepening, begins by sending out standard Gnutella queries in a
sequence. Queries in the sequence differ only in that they have
increasing $ttl$ settings. For example, the first query may be a
broadcast that is $2$ levels deep.  Then the sending client might
wait a pre-specified time for a response, and if no results are
returned, may send out another query with a $ttl$ of $3$.  The
method is therefore parameterized by a sequence of $ttl$ values
and a waiting value.

The method is an improvement over the default protocol when the
queries can be satisfied by nodes closer than the maximum radius
defined by the $ttl$ of the default.  In that case, bandwidth and
processing cost are saved.  Their experiments on a live Gnutella
client showed very good improvements.  The bandwidth used and
processing cost was $19\%$ and $41\%$ of the default policy, and
they argue that the entire network's performance would increase
significantly if each client adopted the iterative deepening
policy.  Some similar results of simulations on different graph
topologies are reported in \cite{lv02ptopsearch}.

\section{Adaptive search }
\label{adaptivesearchsection}
The above sections have examined
strategies for finding a node on a network knowing nothing other
than the identities of one's first and second neighbors. However,
a node can learn about the network over time and adapt its search
strategies. Yang and Garcia-Molina \cite{yang02improvingsearch}
performed experiments on the Gnutella network in which a modified
Gnutella node selectively passed a query onto one of its
neighbors. The neighbors thereafter would follow the standard
Gnutella protocol and broadcast the query to all of their
neighbors. To make the experiment realistic, the queries were
sampled from a collection gathered by passively listening in on
Gnutella traffic.

Yang et al. found that selecting the node which had previously
delivered a specified number of results in the least amount of
time outperformed a strategy which selects a random or a
high-degree neighbor in the first step. The result showed that
adapting the search algorithm to incorporate information learned
about the network can deliver results comparable to BFS
(broadcast) search while using considerably less processing power
and bandwidth.

While nodes can adapt their search strategies based on the
changing performance of nodes in the network, the network itself
can grow and restructure in order to facilitate search. Freenet
\cite{hong01freenet} is an example of a network which dynamically
changes connections and distributes data files as a result of
queries passing through it. Although decentralized, the Freenet
network allows for nodes to specialize in locating subsets of
files and for nodes to direct queries to nodes most likely to be
able to route or satisfy the query.

Each node stores a routing table of files identified by a unique
key and the node which is storing the file. When a node receives a
request for a file listed in its routing table, it forwards the
request to the node listed as having the file. If the there is no
file matching the key, it will forward the request to the location
of a file with the 'closest' key to the key requested. If the
query is eventually satisfied, the file will be passed back along
the same route as the query, and the node will mark the node's
location. In this way nodes learn of the locations of files with
keys similar to the ones already listed in their routing tables
and can specialize in a particular region of the key space,
expediting the search further.

Nodes that reliably answer queries will be added to more routing
tables and hence will be contacted more often than nodes that do
not. In simulations of the network this leads to high degree nodes
acquiring even more connections, and, unsurprisingly, to a
power-law distribution.

\begin{figure}[tbp]
\begin{center}
\includegraphics[scale=0.75]{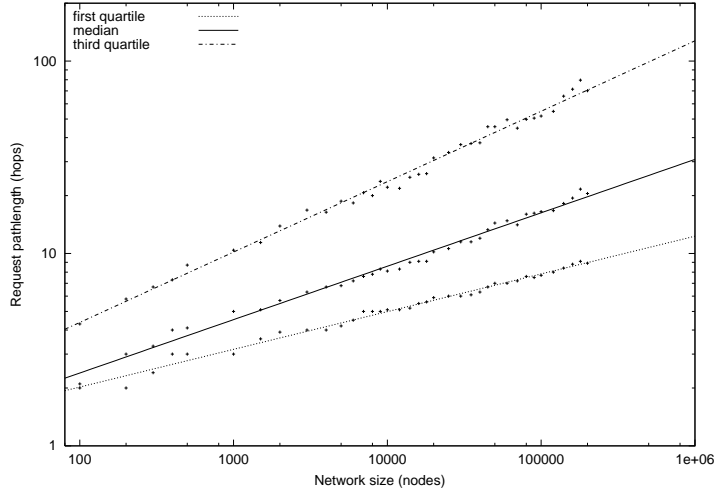}
\end{center}
\caption[Request path length versus Freenet network size]{Request
path length versus Freenet network size. The median path length in
the network scales as $N^{0.28}$. Source: Theodore Hong
\cite{clarke02freenet}} \label{freenetscalability}
\end{figure}

Figure \ref{freenetscalability} shows the number of hops required
to satisfy a request as a simulated Freenet network grows from 20
to 200,000 nodes. The median path length scales as $N^{0.28}$, and
is a mere 8 hops for a network of 10,000 nodes. The result shows
that using a focused search in combination with an adaptive
network can improve scalability of a p2p network.

\section{Conclusion }
\label{conclusionsection}
In this chapter we have shown that local
search strategies in power-law graphs have search costs which
scale sub-linearly with the size of the graph, a fact that makes
them very appealing when dealing with large networks. The most
favorable scaling was obtained by using strategies which
preferentially utilize the high connectivity nodes in these
power-law networks. We also established the utility of these
strategies for searching on the Gnutella peer-to-peer network.
Furthermore, we reviewed the effectiveness of other improvements
to simple broadcast on Gnutella such as iterative deepening and
adaptive search.

Our results on high-degree seeking local search strategies may
extend to social networks.  However, in social networks, it is
clear that people have a wide variety of additional cues to help
them find who and what they need. Nevertheless, our results
suggest that even strategies that neglect those cues may perform
reasonably well on large power-law networks when they take
advantage of the connectedness of nodes.  These strategies have
intuitive appeal, since people naturally ask those they perceive
to be well-connected when trying to locate others in a social
network.

It may not be coincidental that several large networks are
structured in a way that naturally facilitates search. For
example, large social networks, such as the AT\&T call graph
\cite{aiello00massrandom} and the collaboration graph of film
actors, have exponents in the range ($\tau = 2.1 - 2.3 $) which
according to our analysis makes them especially suitable for
searching using simple, local algorithms. Being able to reach
remote nodes by following intermediate links allows communication
systems and people to get to the resources they need and
distribute information within these informal networks.  At the
social level, our analysis supports the hypothesis that highly
connected individuals do a great deal to improve the effectiveness
of social networks in terms of access to relevant resources
\cite{gladwell00tipping}.

Furthermore, it has been shown that the Internet backbone has a
power-law distribution with exponent values between 2.15 and 2.2
\cite{faloutsos99topology}, and web page hyperlinks have an
exponent of 2.1 \cite{barabasi99scaling}. While in the Internet
search is more structured, using routing tables for directing
packets and search engines for finding web pages, high degree
nodes still play a very significant role. Packets are usually
routed through high degree hubs, and people searching for
information on the Web turn to highly connected nodes, such as
directories and search engines, which can bring them to their
desired destinations. On the other hand, a system such as the
power grid of the western United States, which does not serve as a
message passing network, has an exponential degree distribution.

Networks for which locating and distributing information play a
vital role, even without perfect global information, tend to be
power-law with exponents favorable to local search. Actually, we
find it likely that these networks could have evolved so as to
facilitate search and information distribution.

{\large \textbf{Acknowledgements}}

We would like to thank Clip2 for the use of their Gnutella crawl
data.

\bibliography{lada}

\end{document}